\documentclass[12pt,draftclsnofoot,onecolumn]{IEEEtran}
\usepackage{amsfonts}
\usepackage{amsfonts}
\usepackage{setspace}

\usepackage{enumerate}

\usepackage{amsmath}
\usepackage{amssymb}
\usepackage[dvips]{graphicx}
\usepackage{epsfig}
\usepackage{hyperref}

\usepackage{subfigure}

\newcommand{\tsnr}{{\text{\footnotesize{SNR}}}}

\newcommand{\tmax}{\text{{max}}}
\newcommand{\E}{\mathbb{E}}

\newcommand{\Pb}{\bar{P}}

\newcommand{\figsize}{0.45}
\newcommand{\subfigsize}{0.45}

\newcommand{\bR}{{\textbf{R}}}
\newcommand{\bD}{{\textbf{D}}}
\newcommand{\bS}{{\textbf{S}}}

\newcommand{\uutheta}{\breve{\theta}}
\newcommand{\vvtheta}{\overset{\circ}{\theta}}

\newtheorem{Lem1}{Proposition}
\newtheorem{Lem}{Theorem}
\newtheorem{Lemm}{Lemma}
\newtheorem{Rem}{Remark}

\newtheorem{Def}{Definition}

\begin{document}

%
\title{Outage Effective Capacity of Buffer-Aided Diamond Relay Systems Using HARQ with Incremental Redundancy}



%
\author{
\IEEEauthorblockN{Deli Qiao}
\thanks{This work has been supported in part by the National Natural Science Foundation of China (61671205) and the Shanghai Sailing Program (16YF1402600).}
\thanks{Part of this work has been submitted to the 2017 IEEE International Conference on Communications (ICC) \cite{deli-outage}.}
\thanks{D. Qiao is with the School of Information Science\&Technology, East China Normal University, Shanghai, China 200241. Email:
dlqiao@ce.ecnu.edu.cn.}}


\maketitle

\begin{abstract}
In this paper, transmission over buffer-aided diamond relay systems under statistical quality of service (QoS) constraints is studied. The statistical QoS constraints are imposed as limitations on delay violation probabilities. In the absence of channel state information (CSI) at the transmitter, truncated hybrid automatic repeat request-incremental redundancy (HARQ-IR) is incorporated to make better use of the wireless channel and the resources for each communication link. The packets that cannot be successfully received upon the maximum number of transmissions will be removed from buffer, i.e., outage occurs. The \emph{outage effective capacity} of a communication link is defined as the maximum constant arrival rate to the source that can be supported by the \emph{goodput} departure processes, i.e., the departure that can be successfully received by the receiver. Then, the outage effective capacity for the buffer-aided diamond relay system is obtained for HARQ-IR incorporated transmission strategy under the \emph{end-to-end} delay constraints. In comparison with the DF protocol with perfect CSI at the transmitters, it is shown that HARQ-IR can achieve superior performance when the SNR levels at the relay are not so large or when the delay constraints are stringent.
\end{abstract}

\section{Introduction}

In wireless systems, the power of the received signal fluctuates randomly over time due to mobility, changing environment, and multipath fading caused by the constructive and destructive superimposition of the multipath signal components \cite{book}. These random changes in the received signal strength lead to variations in the instantaneous data rates that can be supported by the channel, which may result in transmission errors in deep fading.
Hybrid automatic repeat request (HARQ) protocols have been proposed to enhance the wireless systems performance. Generally, the receiver sends either an acknowledgement (ACK) or negative ACK (NACK) to the transmitter depending on whether the data packet is correctly received or not. The transmitter can decide either to send the next packet or retransmit the same packet upon reception of ACK or NACK, respectively \cite{arq}. The performance of ARQ protocls has been extensively studied in literature (see e.g., \cite{caire-arq}-\cite{jindal-arq} and references therein).

Also, relay channels can be viewed as one of the basic building blocks of wireless systems. Information-theoretic analysis of relay channels has been the research forefront for decades, and has shown the performance improvement in terms of throughput and diversity (see, e.g., \cite{relaycover}-\cite{debbah}). For instance, the authors have considered different relaying strategies in \cite{coopdiver}, and showed that considerable cooperative diversity can be achieved with the relaying schemes. The authors have derived the expressions for the outage probability and throughput for HARQ protocols in relay channels in \cite{alouini}. Of particular interest is the \emph{diamond relay system} in which the communication between a disconnected source and destination is achieved via the help of two or more intermediate relay nodes. The authors have analyzed the capacity bounds for the full-duplex relays with additive white Gaussian channels in \cite{gauss-pr}, while different transmission strategies and achievable rates in half-duplex Gaussian diamond relay channel have been investigated in \cite{hd-diamondr}. The authors have characterized the outage probability and throughput of HARQ protocols with relay selection for the multirelay channels in \cite{debbah}. More recently, buffer-aided relaying in which the relays are equipped with buffers have been shown to further improve the performance of relay systems \cite{schoberrelay}, \cite{diamond-relaysel}. Design and analysis of buffer-aided relay systems have attracted much interest recently \cite{diamond-relaysel}.

Generally, information theoretic analysis do not take into account the buffer/queue limitations. In present wireless systems, diverse quality of service (QoS) requirements are driven by the exponential growth of wireless multimedia traffic that is generated by smartphones, tablets, servers, social networking tools and video sharing sites. In multimedia applications involving e.g., voice over IP (VoIP), streaming video, and interactive video, certain QoS limitations in terms of buffer/delay constraints are imposed so that target levels of performance and quality can be provided to the users. The concept of effective capacity \cite{dapeng} has been incorporated to characterize the maximum constant arrival rate under statistical delay constraints. In case of point-to-point links, there have been some related works investigating the HARQ protocols of wireless channels under statistical QoS constraints recently \cite{qiao-fixed}-\cite{skoglund}. For instance, in \cite{qiao-fixed}, we have analyzed the energy efficiency of fixed rate transmissions under statistical QoS constraints with a simple Type-I HARQ (HARQ-T1) protocol. In this work, we assumed that no outage occurs, i.e., retransmissions are triggered as long as long the receiver does not receive the packet. In \cite{jinho}, the author has analyzed the performance of HARQ with incremental redundancy (HARQ-IR), and showed that with stringent QoS constraints, HARQ-IR can outperform the adaptive transmission system. In \cite{gursoy-harq}, the authors have investigated fixed rate transmissions with HARQ protocols, and obtained the closed-form expression for the effective capacity of HARQ-IR only for loose QoS constraints. In \cite{sami}, the authors have characterized the effective capacity of different HARQ protocols with limited number of transmissions, or deadline of the packets. Outage occurs when the packet is dropped from the buffer while the receiver does not correctly receive the packet. However, the effective capacity obtained does not specify the average throughput that can be correctly received at the receiver. In \cite{skoglund}, the authors have considered the goodput of various HARQ protocols, and proposed a general framework to express effective capacity of HARQ protocols based on a random walk model and recurrence relation formulation. In this paper, we present a study on the buffer-aided diamond relay systems with HARQ-IR under statistical QoS constraints, in the form of limitations on the delay violation probabilities. 

In this work, we assume that the channel state information (CSI) is absent at the transmitters for the links. We first define the \emph{outage effective capacity} as the maximum constant arrival rate that can be supported by the departure processes correctly received at the receiver while satisfying the statistical QoS constraints for a communication link. We show that there is an optimal fixed transmission rate with HARQ-IR scheme. We also demonstrate that the outage effective capacity approaches to the throughput of the link as the delay constraints vanish. We then consider full-duplex decode-and-forward (DF) relays, and assume that the source sends the common information to the relays, which cooperatively deliver the same message to the destination. The relays adopt the Alamouti scheme to enhance the information delivery to the destination. With the proposed HARQ-IR scheme, we derive the outage effective capacity of the buffer-aided diamond relay system and the associated outage probability. For comparison, we also consider the typical DF protocol \cite{superpos-dr} in case of perfect CSI at the transmitter and the receiver for all links, where the common information is sent by the source at the minimum rate of the source-relay links and distributed beamforming is performed at the relays. The contributions of this work can be summarized as follows:
\begin{enumerate}
\item We obtain the outage effective capacity of the goodput processes of a communication link for the HARQ protocols following the spectral radius method, and prove that the limiting behavior of the resulting expression coincides with several well-known results, such as the throughput of HARQ protocols without delay constraints and the effective capacity of HARQ-T1 protocol with unlimited number of transmissions;
\item We propose a HARQ-IR based transmission scheme for the buffer-aided diamond relay systems with perfect CSI at the receiver only for each link, and characterize the outage effective capacity of the proposed scheme under the statistical delay constraints;
\item Through numerical evaluations, we demonstrate the superiority of the proposed scheme with respect to the DF protocol with perfect CSI at the transmitter and receiver of each link when the $\tsnr$ at the relay is relatively small or when the delay constraints are relatively stringent.
    \end{enumerate}

The rest of this paper is organized as follows. Section II introduces preliminaries on the diamond relay channel model, and reviews the HARQ-IR operations. In Section III, we briefly discuss the statistical delay constraints and define the outage effective capacity for one-hop links. Section IV discusses the effective capacity analysis method for two-hop links, and characterize the outage effective capacity of the buffer-aided diamond relay systems. Numerical results are provided in Section V. Finally, Section VI concludes this paper.

\section{Preliminaries}

\subsection{System Model}

\begin{figure}
\begin{center}
\includegraphics[width=\subfigsize\textwidth]{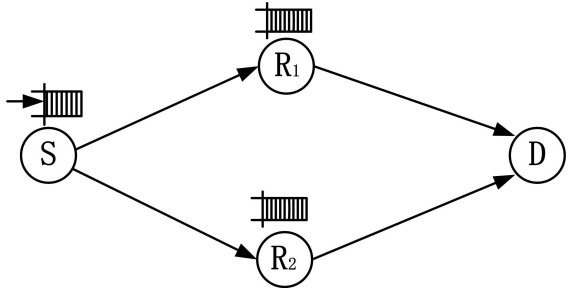}
\caption{The diamond-relay system model.}\label{fig:systemmodel-dr}
\end{center}
\end{figure}

We consider a buffer-aided diamond relay communication link as depicted in Fig.
\ref{fig:systemmodel-dr}. The source sends information to the destination via the help of two parallel relays. We assume that there is no direct link between the source and the destination. Also, there is no link between the relays. In this model, there are buffers of infinite size at both the source and relays. In this work, we assume full-duplex relay such that transmission and reception can be performed simultaneously. 

The discrete-time input and output relationships in the $i$th symbol duration are given by
\begin{align}
Y_{r_j}[i]&=g_{sr_j}[i]X_s[i]+n_{r_j}[i], \,j=1,2,\\
Y_d[i]&=g_{r_1d}[i]X_{r_1}[i]+g_{r_2d}[i]X_{r_2}[i]+n_d[i],
\end{align}
where $X_{k}$ for $k\in\{s, r_{1},r_2\}$ denote the input signal from the source $\bS$ and the relay $\bR_{j},j=1,2$, respectively. The inputs are subject to individual average
energy constraints $\E\{|X_k|^2\}\le \Pb_k/B, k\in\{s,r_1,r_2\}$, where $B$ is the bandwidth. $Y_{r_j}, Y_d$ represent the received signal at the relay $\bR_j$ and the destination $\bD$, respectively. We assume that the fading coefficients $g_{sr_j}, g_{r_jd}$ are jointly stationary and ergodic discrete-time processes, and we denote the magnitude-square of the fading
coefficients by $z_{sr_j}[i]=|g_{sr_j}[i]|^2$ and $z_{r_jd}[i]=|g_{r_jd}[i]|^2$. Denote $\mathbf{z} = (z_{sr_1},z_{sr_2},z_{r_1d},z_{r_2d})$. Assuming that there are $B$
complex symbols per second, we can easily see that the symbol energy
constraint of $\Pb_k/B$ implies that the channel input has a power
constraint of $\Pb_k$. Above, in the channel input-output
relationships, the noise component $n_k[i]$ is a zero-mean,
circularly symmetric, complex Gaussian random variable with variance
$\E\{|n_k[i]|^2\} = N_0$ for $k \in\{r_1,r_2,d\}$. The additive Gaussian noise
samples $\{n_k[i]\}$ are assumed to form an independent and
identically distributed (i.i.d.) sequence. We denote the
signal-to-noise ratio at source as $\tsnr_s=\frac{\Pb_s}{N_0 B}$, and at relays as $\tsnr_{r_j} = \frac{\Pb_{r_j}}{N_0 B}, j=1,2$.

\subsection{HARQ-IR}\label{sec:harqmodel}

Consider a link composed of one transmitter and one receiver under block fading in which the fading stays constant for a block of $T$ seconds and changes independently from one block to another. We assume that a packet of $L$ bits is intended to be transmitted over the wireless channel in each frame. Specifically, after each successful transmission, the transmitter attempts to send $L$ bits in the next frame. So the fixed transmission rate is termed as $L$ bits/block. We assume that upon successful reception of the packet, the receiver sends an ACK to the transmitter, and the packet can be removed from the buffer. If a decoding failure occurs, the receiver sends a NACK to the transmitter and requests another round of retransmissions for the packet if the maximum number of transmissions for the packet is not reached. On the other hand, when the maximum number of transmissions for the packet is reached, the packet will be removed from the buffer without the need of ACK or NACK. Therefore, outage occurs at the maximum round of transmission, i.e., $M^{\text{th}}$ transmission, if the packet is discarded from the buffer while the receiver does not correctly receive this packet.
\begin{figure}
\begin{center}
\includegraphics[width=\figsize\textwidth]{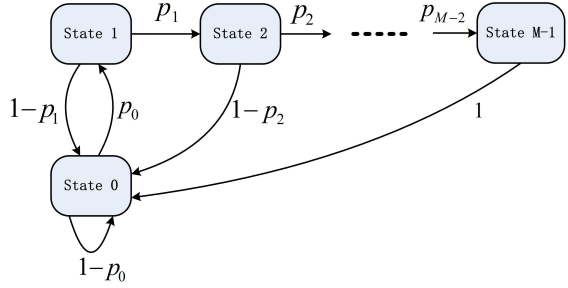}
\caption{The state transition model.}\label{fig:state}
\end{center}
\end{figure}

We can model the buffer activity at the end of each frame as a discrete-time Markov process \cite{sami}. Fig. \ref{fig:state} depicts the state transition model. State 0 denotes that the packet is removed from the buffer, and state $m$ represents the number of retransmissions for the packet, where no packet is removed from the buffer. Define $p_m$ as the decoding failure probability at the $m^{\text{th}}$ retransmission such that the system enters State $m+1$ with probability $p_m$, while the system enters state 0 with probability $1-p_m$. On the other hand, regardless of the decoding result at the end of $(M-1)^\text{th}$ retransmission, the system goes to State 0
with probability $1=p_{M-1} + (1-p_{M-1})$ since the maximum number of transmissions is reached and the packet is removed immediately from the buffer. Then, the state transition matrix is given by
\begin{align}\label{eq:stateprob}
\mathbf{P} = \left[
\begin{array}{lllll}
1-p_0 & 1-p_1  & \cdots & 1-p_{M-2} & 1\\
p_0 & 0  & \cdots & 0 & 0\\
0 & p_1  &\cdots & 0 & 0\\
\vdots & \vdots  & \ddots & \vdots & \vdots\\
0 & 0  & \cdots & p_{M-2} & 0
\end{array}
\right]
\end{align}
where $P_{ij}$ denotes the probability of state transition from State $j$ to State $i$.

In the HARQ-IR protocol, the transmitter encodes the packet according to a codebook of length $MTB$, and the codewords are divided into subblocks of the same length with $TB$ symbols. During each frame, only one subblock is sent to the receiver, and the receiver decodes the message using the current subblock combined with the previously received sublocks of the packet. Then, we know that the receiver can successfully decode the packet after the $m^{\text{th}}$ ($0\le m\le M-1$) retransmission only if the following condition is satisfied \cite{caire-arq}
\begin{align}
L\le \sum_{i=0}^mTB\log_2\left(1+\tsnr z_i\right).
\end{align}
We can express the state transition probabilities as $p_0 = \Pr\left\{z<\frac{2^{L/TB}-1}{\tsnr}\right\}$, and for $m=1,\ldots,M-1$,
\begin{align}\label{eq:tpir}
p_{m}
& = \frac{\Pr\left\{L> \sum_{i=0}^mTB\log_2\left(1+\tsnr z_i\right)\right\}}{\Pr\left\{L >  \sum_{i=0}^{m-1}TB\log_2\left(1+\tsnr z_i\right)\right\}}.
\end{align}

Define the outage probability after $m$-th transmission rounds as
\begin{align}
P_{\text{out},m} = \Pr\left\{\sum_{i=0}^{m-1}TB\log_2\left(1+\tsnr z_i\right)<L\right\}.
\end{align}
In the absence of delay constraints, the throughput of truncated HARQ, i.e., \emph{goodput}, is known to be \cite{caire-arq}
\begin{align}
R_{\text{HARQ}} = \frac{L}{TB}\frac{(1-P_{\text{out},M})}{\sum_{m=0}^{M-1}P_{\text{out},m}},\,\,\text{bps/Hz},
\end{align}
where $P_{\text{out},0} = 1$.

\section{Effective Capacity Analysis in One-Hop Links}
In this section, we first review the preliminaries on the statistical delay constraints, and then obtain the outage effective capacity for a communication link with the parameters discussed above.

\subsection{Statistical Delay Constraints for One-Hop Links}
Suppose that the queue is stable and that both the arrival process
$a[n]$ and service process $c[n]$ satisfy the G\"{a}rtner-Ellis limit, i.e., for all $\theta\ge0$, there exists a differentiable logarithmic moment generating function
(LMGF) $\Lambda_A(\theta)$ such that\footnote{Throughout the text,
logarithm expressed without a base, i.e., $\log(\cdot)$, refers to
the natural logarithm $\log_e(\cdot)$.}
$\lim_{n\to\infty}\frac{\log \E\{e^{\theta\sum_{i=1}^n
a[n]}\}}{n}=\Lambda_A(\theta)$,
and a differentiable LMGF $\Lambda_C(\theta)$ such that
$\lim_{n\to\infty}\frac{\log \E\{e^{\theta\sum_{i=1}^n
c[n]}\}}{n}=\Lambda_C(\theta)$.
If there exists a unique $\theta^*>0$ such that
\begin{align}\label{eq:queuestable}
\Lambda_A(\theta^*)+\Lambda_C(-\theta^*)=0,
\end{align}
then \cite{changbook}
\begin{align} \label{eq:QoSexponentdef}
\lim_{Q_{\tmax}\to\infty}\frac{\log
\Pr\{Q>Q_{\tmax}\}}{Q_{\tmax}}=-\theta^*.
\end{align}
where $Q$ is the stationary queue length.

For large $Q_{\tmax}$, we have the approximation for the buffer violation probability: $\Pr\{Q>Q_{\tmax}\}\approx e^{-\theta^* Q_{\tmax}}$. Hence, while larger $\theta$ corresponds to stricter queueing constraints, smaller $\theta$ implies looser queueing constraints. Then, equivalently, we have the queueing delay violation probability as $\Pr\{D>D_{\tmax}\} \approx  e^{-J(\theta) D_\tmax}$, where $$J(\theta) = -\Lambda_C(-\theta)$$ is the statistical delay exponent associated with the queue, with $\Lambda_C(\theta)$ the LMGF of the service rate. Then, the maximum constant arrival rate to the queue for given $\theta>0$ is expressed as
\begin{align}
R_E(\theta)=-\frac{\Lambda_C(-\theta)}{\theta TB},\,\,\text{bps/Hz}.
\end{align}
%

\subsection{Outage Effective Capacity}\label{sec:rate}

While the authors in \cite{sami} considered the departure processes of the source queue, we focus on the \emph{goodput} departure processes that can be correctly received at the receiver similar to \cite{skoglund}. According to (\ref{eq:queuestable}), we define the outage effective capacity as the maximum constant arrival rate to the source that can be supported by the \emph{goodput} processes. Then, we can obtain the following result.
\begin{Lem}\label{theo:ec}
For the fixed rate transmissions with HARQ protocols, given QoS exponent $\theta>0$, $\tsnr>0$, and maximum number of transmissions $M$, the outage effective capacity is given by
\begin{align}
&R_{\text{out}}(\tsnr,\theta) 
= \frac{1}{TB}\max_{L\ge0}\left\{-\frac{\Lambda(-\theta)}{\theta}\right\}\\
& = \max_{L\ge0}\left\{-\frac{1}{\theta TB}\log\left(p_0\left(P_{\text{out}} + (1-P_{\text{out}}) e^{-\theta L}\right)y^*\right)\right\}\label{eq:ecfixedL}\\
& = -\frac{1}{\theta TB}\log\left(p_{0,\text{opt}}\left(P_{\text{out,opt}} + (1-P_{\text{out,opt}}) e^{-\theta L_{\text{opt}}}\right)y_{\text{opt}}^*\right)\label{eq:ecoptL}
\end{align}
where $L_{\text{opt}}$ is the optimal finite fixed transmission rate that solves (\ref{eq:ecfixedL}), $y^*$ is the only unique real positive root of $f(y)=0$ with
\begin{align}\label{eq:fy}
f(y) &= y^M - \frac{1-p_0}{p_0}y^{M-1}
-\sum_{m=1}^{M-2}\frac{(1-p_m)p_{m-1}\cdots p_1}{p_0^{m} \left(P_{\text{out}} + (1-P_{\text{out}}) e^{-\theta L}\right)^m}y^{M-1-m}\nonumber\\
&\hspace{2cm}
-\frac{p_{M-2}\cdots p_1}{p_0^{M-1} \left(P_{\text{out}} + (1-P_{\text{out}}) e^{-\theta L}\right)^{M-1}}
\end{align}
for given $L$, and $y^*_{\text{opt}}$ is the only unique real positive root of $f(y)=0$ with $L=L_{\text{opt}}$. The outage probability associated can be expressed as
\begin{align*}
P_{\text{out,opt}} = \prod_{m=0}^{M-1}p_{m,\text{opt}}=\Pr\left\{\sum_{m=0}^{M-1}TB\log_2(1+\tsnr z_m) < L_{\text{opt}}\right\}
\end{align*}
where $p_{m,\text{opt}}$ denotes the state transition probability obtained with $L_{\text{opt}}$.
\end{Lem}
\emph{Proof:} See Appendix \ref{app:ec}.\hfill$\square$

\begin{Rem}\label{rem:optL}
Above, we did not specify how $L_{\text{opt}}$ is obtained. Since there is no closed form expression for $y^*$ which depends on $L$ nonlinearly, we can only solve (\ref{eq:ecfixedL}) numerically. For instance, in the following numerical results, we employ branch-and-bound method to find $L_{\text{opt}}$. In general, $L_{\text{opt}}$ depends on $\theta$, $\tsnr$, and $M$. Note that the rate expression in (\ref{eq:ecfixedL}) is applicable for all $\theta>0$, in stark difference from the results in \cite{gursoy-harq}, where the closed-form expression of effective capacity is obtained for small $\theta$. Also, we characterized the outage probability that was not treated in \cite{gursoy-harq}. Note also that the rate expression in (\ref{eq:ecoptL}) is different from the results in \cite{sami}, where packet drop is not considered, and the results in \cite{skoglund}, where the results are in matrix form based on a random walk model and recurrence relation formulation.
\end{Rem}

In the absence of statistical QoS constraints, we have the following result.
\begin{Lem1}\label{theo:eclimit}
As $\theta\to0$, the outage effective capacity with HARQ protocols is given by
\begin{align}
\lim_{\theta\to0}R_{\text{out}}(\tsnr,\theta) &= \frac{L_{\text{opt}}}{TB}\frac{(1-P_{\text{out,opt},M})}{\sum_{m=0}^{M-1}P_{\text{out,opt},m}},\,\,\text{bps/Hz}.
\end{align}
\end{Lem1}
\emph{Proof:} See Appendix \ref{app:eclimit}.\hfill$\square$

\begin{Rem}
Note that the outage effective capacity approaches to the maximum goodput of the HARQ protocols, i.e., $\max_{L\ge0}R_{\text{HARQ}}$, as the statistical QoS constraints vanish.
\end{Rem}

\begin{Rem}
The results in Theorem \ref{theo:ec} is generic, and can be applied to other HARQ protocols as well, e.g., HARQ-T1, where the transmitter sends the same packet in each frame during retransmissions and the receiver decodes the packet successfully if the instantaneous channel rate is greater than the transmission rate, and HARQ-chase combining (HARQ-CC), where the receiver can make use of the received signals in the previous frames through maximum ratio combining. Note that it has been verified that HARQ-IR performs better than the other schemes under statistical delay constraints \cite{gursoy-harq}. As $M\to\infty$, the outage probability vanishes and the outage effective capacity is exactly the constant arrival rate supported by the departure processes. Here, we give an example of HARQ-T1 when $M\to\infty$.
\begin{Lem1}\label{theo:t1minf}
For the fixed rate transmissions with HARQ-T1 protocol, the outage effective capacity for a given QoS exponent $\theta>0$ and $\tsnr>0$ approaches to the following value as $M\to\infty$:
\begin{align}\label{eq:t1minfrate}
&\lim_{M\to\infty}R_{\text{out}}(\tsnr,\theta)
 = \max_{L\ge0}\Bigg\{-\frac{1}{\theta TB}\log\bigg(1 - \Pr\left\{z>\frac{2^{L/TB}-1}{\tsnr}\right\}
\left(1-e^{-\theta L}\right)\bigg)\Bigg\}.
\end{align}
\end{Lem1}
\emph{Proof:} See Appendix \ref{app:t1minf}.\hfill$\square$

Obviously, (\ref{eq:t1minfrate}) coincides with the result in \cite[(11)]{qiao-fixed}. 
\end{Rem}

\section{Outage Effective Capacity in Buffer-Aided Diamond Relay Systems}\label{sec:onehop}

In this section, we first briefly discuss the statistical delay constraints for two-hop links, and then define and characterize the outage effective capacity for the buffer-aided diamond relay systems under consideration.

\subsection{Statistical Delay Constraints for Two-Hop Links}
In this work, we seek to identify the maximum constant arrival rate to the source that can be supported by the \emph{goodput} processes successively received at the destination of the diamond relay system using HARQ-IR while satisfying the statistical delay constraints. Therefore, we need to guarantee that the data transmission of all information flows should satisfy the statistical delay constraints. Since there is \emph{no link} between the relays, we have at most two concatenated queues for the information flow. Consider two concatenated queues with statistical queueing constraints specified by $\theta_1$ and $\theta_2$, for queue 1 and queue 2, respectively. Given the queueing constraints specified by $\theta_1$ and $\theta_2$ with (\ref{eq:QoSexponentdef}) satisfied for each queue, we define
\begin{align}\label{eq:J1J2eq}
J_1(\theta_1)=-\Lambda_{C,1}(-\theta_1),\,\,\text{and}\,\, J_2(\theta_2)=-\Lambda_{C,2}(-\theta_2),
\end{align}
where $\Lambda_{C,1}(\theta_1)$ and $\Lambda_{C,2}(\theta_1)$ are the LMGF functions of the service rate of queue 1, 2, respectively. For data going through both queues, the end-to-end queueing delay violation probability can be characterized as
\begin{align}
\Pr\{D_1+D_2>D_\tmax\} 
&
\doteq 1 - \int_0^{D_\tmax} \int_{0}^{D_\tmax - D_1}p_{D}(D_1)p_D(D_2)dD_2dD_1\nonumber\\
& = \left\{
\begin{array}{ll}
\frac{J_1(\theta_1)e^{-J_2(\theta_2)D_{\tmax}}-J_2(\theta_2)e^{-J_1(\theta_1)D_{\tmax}}}{J_1(\theta_1)-J_2(\theta_2)},& J_1(\theta_1)\neq J_2(\theta_2)\\
\left(1+J_1(\theta_1)D_{\tmax}\right)e^{-J_1(\theta_1)D_{\tmax}},&J_1(\theta_1)=J_2(\theta_2).
\end{array}\right.\label{eq:delayprob}
\end{align}
Thereby, we need to guarantee that
\begin{align}\label{eq:queue12cond}
\Pr\{D_1+D_2>D_{\tmax}\}\le \varepsilon.
\end{align}
In this way, we can guarantee that the data transmissions through the relays, i.e., information flows over two queues at the source and the relays, satisfy the statistical delay constraints. Then, the delay constraints of the whole system can be satisfied. Note that $(\varepsilon,D_\tmax)$ characterizes the statistical delay constraints with maximum delay violation probability $\varepsilon$ and maximum delay $D_\tmax$. 
To facilitate the following analysis, we need the following tradeoff between the delay exponents of any concatenated two queues, i.e., $J_1(\theta_1)$ and $J_2(\theta_2)$.
\begin{Lemm}[\cite{deli-twohopend}]\label{lemm:J1J2relation}
Consider the following function
\begin{align}\label{eq:J1J2relation}
\vartheta(J_1(\theta_1),J_2(\theta_2))&=\frac{J_2(\theta_2)e^{-J_1(\theta_1)D_\tmax} - J_1(\theta_1)e^{-J_2(\theta_2)D_\tmax}}{J_2(\theta_2)-J_1(\theta_1)} 
= e^{-J_0 D_\tmax}=\varepsilon, \,\text{for} \, 0\le\varepsilon\le1,
\end{align}
where $J_0=-\frac{\log(\varepsilon)}{D_\tmax}$ is defined as the statistical delay exponent associated with $(\varepsilon,D_\tmax)$. Denoting $J_2(\theta_2) = \Phi(J_1(\theta_1))$ as a function of $J_1(\theta_1)$, we have
\begin{enumerate}[a)]

\item $\Phi$ is continuous. For $J_1(\theta_1)=J_{th}(\varepsilon)$, we have
\begin{align}
\Phi(J_1(\theta_1)) =  J_{th}(\varepsilon),
\end{align}
where
\begin{align}\label{eq:Jfunctioncond}
J_{th}(\varepsilon) = -\frac{1}{D_{\tmax}}\left(1+\mathcal{W}_{-1}\left(-\frac{\varepsilon}{e}\right)\right),
\end{align}
where $\mathcal{W}_{-1}(\cdot)$ is the Lambert W function, which is the inverse function of $y=xe^x$ in the range $(-\infty,-1]$.

\item $\Phi$ is strictly decreasing in $J_1(\theta_1)$.

\item $\Phi$ is convex in $J_1(\theta_1)$.

\item $J_1(\theta_1)\in[J_0,\infty)$, and $J_2(\theta_2)=\Phi(J_1(\theta_1))\in[J_0,\infty)$.

\end{enumerate}
\end{Lemm}

\subsection{Effective Capacity Analysis of Diamond-Relay Systems}

If we define $\theta_1$, $\theta_{r_1}$ and $\theta_{r_2}$ as the statistical queueing constraints at the source and the relays, respectively. For different information flows over the relays, we will have different two-hop channels with queueing constraints $(\theta_1,\theta_{r_1})$ and $(\theta_1,\theta_{r_2})$, respectively.  Assume that the equivalent constant arrival rate at the source is $R\ge0$. Consider any realization $(\theta_1,\theta_2)$ of any two concatenated queues. Denote $\Omega$ as the set of pairs $(\theta_1,\theta_2)$ such that (\ref{eq:queue12cond}) can be satisfied. To satisfy the queueing constraint at queue 1, i.e., queue at the source, we should have
$\tilde{\theta}\ge\theta_1$,
where $\tilde{\theta}$ is the solution to
\begin{align}\label{eq:queue1cond}
R = -\frac{\Lambda_{C,1}(-\tilde{\theta})}{\tilde{\theta}},
\end{align}
and $\Lambda_{C,1}(\theta)$ is the LMGF of the \emph{goodput} service for queue 1, i.e., service processes successively received at queue 2 (any relay).

Also, in order to satisfy the queueing constraint of queue 2, we must have
$\hat{\theta}\ge\theta_2$,
where $\hat{\theta}$ is the solution to
\begin{align}\label{eq:queue2cond}
\Lambda_{A,2}(\theta)+\Lambda_{C,2}(-\theta)=0.
\end{align}
where $\Lambda_{A,2}(\theta)$ is the LMGF of the \emph{goodput} arrivals at queue 2, $\Lambda_{C,2}(\theta)$ is the LMGF of the \emph{goodput} service of queue 2, i.e., service processes successively received at destination. Note that we need to consider the queues at the relays together with the queue at the source.

Denote $\Omega$ as the set of pairs $(\theta_1,\theta_2)$ of two concatenated buffers such that (\ref{eq:queue12cond}) can be satisfied. Now, \emph{outage effective capacity} of the buffer-aided diamond relay system under statistical delay constraints $(\varepsilon,D_\tmax)$ can be formulated as follows.
\begin{Def}\label{def:ecdef}
The outage effective capacity of the buffer-aided diamond relay system with statistical delay constraints specified by $(\varepsilon,D_\tmax)$ is given by
\begin{align}\label{eq:effdefi}
R(\varepsilon,D_\tmax)=\sup_{(\theta_1,\theta_{r_1})\in\Omega,(\theta_1,\theta_{r_2})\in\Omega} R.
\end{align}
Hence, outage effective capacity is now the maximum constant arrival rate that can be supported by the \emph{goodput} processes successfully received at the destination of the diamond relay system under statistical delay constraints.
\end{Def}

%

\subsection{Outage Effective Capacity of Diamon-Relay Links with HARQ-IR}
In this part, we study the performance of HARQ-IR in the buffer-aided diamond-relay channels. We assume that common messages are sent to the relays and the relays cooperate in the information delivery to the destination such that the queue dynamics at the relays are the same. We consider the \emph{end-to-end} delay constraints, and identify the maximum constant arrival goodput to the source and the end-to-end outage probability while satisfying the statistical delay constraints.

\subsubsection{Decode-and-Forward (DF)}\label{sec:ratedf}
As a comparison, we consider the decode-and-forward (DF) scheme \cite{superpos-dr}, in which case the CSI is also available at the transmitter for each link and each relay must successfully decode the common message transmitted by the source node, and later the relays can cooperatively beamform their transmissions to the destination. We assume that the transmission power levels at the source and relays are fixed and hence no power control is employed (i.e., nodes are subject to short-term power constraints). We further assume that the channel capacity for each link can be achieved, i.e., the service processes are equal to the instantaneous Shannon capacities of the links such that there is no decoding error. Then, the service rate leaving the queue at the source is given by
\begin{align}\label{eq:ratesdf}
C_{s} = TB\log_2(1+\tsnr_s \min\{z_{sr_1},z_{sr_2}\}).
\end{align}
Also, the rates leaving the queues at the relays are the same, and are given by
\begin{align}\label{eq:raterdf}
\hspace{-.7cm}C_{r_1} = C_{r_2} &= TB\log_2\left(1+\left(\sqrt{\tsnr_{r_1} z_{r_1d}} + \sqrt{\tsnr_{r_2} z_{r_2d}}\right)^2 \right).
\end{align}

Above, the rates are given in terms of bits/block. Note that the arrival rates and departure rates of the queues at the relays are always the same, and hence the queueing activities have the same pattern. Therefore, the system simplifies to the two-hop channel. Then, we can obtain the effective capacity similar to the discussions in \cite{deli-twohopend}. In this scheme, the end-to-end outage probability is zero, i.e., all departure processes can be successfully received at the destination.

\subsubsection{HARQ-IR}\label{sec:harqdr}

We assume perfect CSI is available only at the receiver for each link, in which case HARQ-IR is incorporated for the transmissions. Similar to the discussion in Section \ref{sec:harqmodel}, we first assume that a packet of $L$ bits is intended to be transmitted in each frame for the each hop and obtain the outage effective capacity associated with $L$. Then, we optimize over $L\ge0$ to find the optimal $L_{\text{opt}}$ that leads to the maximum outage effective capacity.

The operations of HARQ-IR can be described as follows:
\begin{enumerate}[a)]
\item In the first hop, the source tries to send the same information to the relays. Note that only after reception of ACKs from all relays, the packet can be removed from the buffer, and the source attempts to send $L$ bits in the next frame. Again, we model the source buffer activity at the end of each frame as a discrete-time Markov process. Define $p_{s,m}$ as the decoding failure probability at the $m^{\text{th}}$ retransmission such that the system enters State $m+1$ with probability $p_{s,m}$, while the system enters state 0 with probability $1-p_{s,m}$. On the other hand, regardless of the decoding result at the end of $(M-1)^\text{th}$ retransmission, the system goes to State 0 with probability $1=p_{s,M-1} + (1-p_{s,M-1})$ since the maximum number of transmissions is reached and the packet is removed immediately from the source buffer. The state transition matrix $\mathbf{P}_s$ can be expressed similar to (\ref{eq:stateprob}) with values $p_{s,m}$ instead. For each relay, we know that the relay can successfully decode the packet after the $m^{\text{th}}$ ($0\le m\le M-1$) retransmission only if the following condition is satisfied
    \begin{align}
    L\le \sum_{i=0}^mTB\log_2\left(1+\tsnr z_{sr_j,i}\right),j=1,2.
    \end{align}
    Therefore, we can express the state transition probabilities as
    \begin{align}
    p_{s,0} &= \Pr\left\{\left\{z_{sr_1}<\frac{2^{L/TB}-1}{\tsnr}\right\}\bigcup \left\{z_{sr_2}<\frac{2^{L/TB}-1}{\tsnr}\right\}\right\}\nonumber\\
    & = 1 - \Pr\left\{z_{sr_1}\ge\frac{2^{L/TB}-1}{\tsnr}\right\}\Pr\left\{z_{sr_2}\ge\frac{2^{L/TB}-1}{\tsnr}\right\}
    \end{align}
    and for $m=1,\ldots,M-1$, we have 
    \begin{align}
    p_{s,m}
    & = \frac{\Pr\left\{\{L> \sum_{i=0}^mTB\log_2\left(1+\tsnr z_{sr_1,i}\right)\}\bigcup \{L> \sum_{i=0}^mTB\log_2\left(1+\tsnr z_{sr_2,i}\right)\}\right\}}{\Pr\left\{\{L >  \sum_{i=0}^{m-1}TB\log_2\left(1+\tsnr z_{sr_1,i}\right)\} \bigcup \{L> \sum_{i=0}^{m-1}TB\log_2\left(1+\tsnr z_{sr_2,i}\right)\}\right\}}\label{eq:tpir-dr1}\\
    & = \frac{1- \Pr\left\{L \le \sum_{i=0}^mTB\log_2\left(1+\tsnr z_{sr_1,i}\right)\right\} \Pr\left\{ L\le \sum_{i=0}^mTB\log_2\left(1+\tsnr z_{sr_2,i}\right)\right\}}{1- \Pr\left\{L \le \sum_{i=0}^{m-1}TB\log_2\left(1+\tsnr z_{sr_1,i}\right)\right\} \Pr\left\{ L\le \sum_{i=0}^{m-1}TB\log_2\left(1+\tsnr z_{sr_2,i}\right)\right\}}.\label{eq:tpir-dr2}
    \end{align}

\item In the second hop, the relays attempt to send the same message to the destination. Following the idea of treating the relays as distributed antennas, we can adopt the Alamouti scheme to improve the achievable rate. Specifically, we divide the frame into two slots of equal length $TB/2$. In one slot, the relay $\bR_1$ sends message $x_1$, and the relay $\bR_2$ sends message $x_2$. In the other slot, the relay $\bR_1$ sends message $x_2^*$, and the relay $\bR_2$ sends message $-x_1^*$. Then, the achievable rate for the second hop in each frame can be expressed as
    \begin{align}\label{eq:rdrate}
    R = TB\log_2\left(1+\tsnr_{r_1} z_{r_1d} + \tsnr_{r_2} z_{r_2d} \right),\,\,\text{bits/block}.
    \end{align}
    Note that the arrival rates and departure rates of the queues at the relays are always the same, and hence the queueing activities have the same pattern. Now, for the Makov process associated with the buffer activities at the relays, we have the state transition matrix $\mathbf{P}_r$ with state transition probabilities as $p_{r,0} = \Pr\left\{\tsnr_{r_1} z_{r_1d} + \tsnr_{r_2} z_{r_2d}<2^{L/TB}-1\right\}$, and for $m=1,\ldots,M-1$,
    \begin{align*}
    p_{r,m}
    & = \frac{\Pr\left\{L> \sum_{i=0}^mTB\log_2\left(1+\tsnr_{r_1} z_{r_1d,i} + \tsnr_{r_2} z_{r_2d,i}\right)\right\}}{\Pr\left\{L >  \sum_{i=0}^{m-1}TB\log_2\left(1+\tsnr_{r_1} z_{r_1d,i} + \tsnr_{r_2} z_{r_2d,i}\right)\right\}}.
    \end{align*}

\end{enumerate}

We can obtain the statistical delay exponent for each hop as
\begin{align}
J_1(\theta_1) = -\Lambda_{C,1}(-\theta_1) = -\log sp\left\{\mathbf{P}_s\phi_s(-\theta_1)\right\}\\
J_2(\theta_2) = -\Lambda_{C,2}(-\theta_2) = -\log sp\left\{\mathbf{P}_r\phi_r(-\theta_2)\right\}
\end{align}
where $\phi_s(\theta_1)=diag(P_{\text{out},s} + (1 - P_{\text{out},s}) e^{\theta L},\underbrace{1,\ldots,1}_{M-1})$ and $\phi_r(\theta_2)=diag(P_{\text{out},r} + (1 - P_{\text{out},r}) e^{\theta L},\underbrace{1,\ldots,1}_{M-1})$ are diagonal matrices with each component given by the moment generating functions of the \emph{goodput} processes in $M$ states of the Markov processes $\mathbf{P}_s$ and $\mathbf{P}_r$, where $P_{\text{out},k}=\prod_{m=0}^{M-1}p_{k,m},k=s,r$ denotes the outage probability of the first and second hop, respectively.

Given $L>0$, we denote $J_{1,\tmax} = \lim_{\theta_1\to\infty}J_1(\theta_1)$ and $J_{2,\tmax} = \lim_{\theta_2\to\infty}J_2(\theta_2)$ as the maximum delay exponent of the first and second hop, which is obtained as the statistical queueing constraints approach infinity. We can show the following results.
\begin{Lem1}
With the HARQ-IR protocol, $J_{1,\tmax} $ and $J_{2,\tmax} $ are finite if $p_{k,0}\neq0,k=s,r$.
\end{Lem1}
\emph{Proof:} First, it can be easily verified that $P_{\text{out},k}\neq0,k=s,r$ if $p_{k,0}\neq0,k=s,r$ since $p_{k,m}\neq0,k=s,r,m=1,\ldots,M-1$. As $\theta\to\infty$, we can see from (\ref{eq:fy}) that $y^*$ will be the solution to the following equation
\begin{align}
\lim_{\theta\to\infty}f(y) &= y^M - \frac{1-p_0}{p_0}y^{M-1}
-\sum_{m=1}^{M-2}\frac{(1-p_m)p_{m-1}\cdots p_1}{p_0^{m}P^m_{\text{out}}} y^{M-1-m}
-\frac{p_{M-2}\cdots p_1}{p_0^{M-1}P_{\text{out}}} = 0.
\end{align}
Obviously, $y^*$ approaches to some finite value. Hence, $\lim_{\theta\to\infty}J(\theta)=\lim_{\theta\to\infty}-\Lambda(-\theta)=-\log\left(p_{0}P_{\text{out}}y^*\right)$ is finite, which implies that $J_{1,\tmax} $ and $J_{2,\tmax} $ are finite. \hfill$\square$

\begin{Rem}\label{rem:jmax}
Note that $p_0\neq 0 $ means that the possibility of failure to decode the packet in the first transmission is not zero. For the fading distributions such as Rayleigh and Nakagami-m, we can see that $p_0 = \Pr\left\{z<\frac{2^{L/TB}-1}{\tsnr}\right\}$ is greater than zero for all $L>0$.
\end{Rem}

Define
\begin{align*}
\Omega_{\varepsilon} = \{(\theta_1,\theta_2): \text{ $J_1(\theta_1)$ and $J_2(\theta_2)$ are solutions to }\,(\ref{eq:queue12cond}) \,\text{w/ equality}\}.
\end{align*}
With the above characterizations, we can obtain the following results.
\begin{Lem}\label{theo:harqresult}
Given $L>0$, the outage effective capacity of the buffer-aided diamond relay systems with HARQ-IR strategy subject to statistical delay constraints specified by $(\varepsilon,D_{\tmax})$ is given by the following:

\textbf{\underline{Case I}}: If $\vartheta(J_{1,\tmax},J_{2,\tmax}) > \varepsilon$,
\begin{gather}
\hspace{-.5cm}R_{HARQ-IR}(\varepsilon,D_\tmax,L)=0,
\end{gather}

\textbf{\underline{Case II}}: Otherwise,

\textbf{\underline{Case II.a}}: If $J_{1,\tmax}<J_{th}(\varepsilon)$,
\begin{gather}
\hspace{-.3cm}R_{HARQ-IR}(\varepsilon,D_\tmax,L)= 
\frac{J_{1}(\vvtheta_1)}{\vvtheta_{1}},
\end{gather}
where 
$\vvtheta_1$ is the smallest value of $\theta_1$ with $(\theta_1,\theta_{2})\in\Omega_{\varepsilon}$ satisfying
\begin{align}
J_1(\theta_1) = J_2(\theta_2)+J_1(\theta_1-\theta_2).
\end{align}

\textbf{\underline{Case II.b}}: If $J_{2,\tmax}<J_{th}(\varepsilon)$,
\begin{gather}
\hspace{-.5cm}R_{HARQ-IR}(\varepsilon,D_\tmax,L)=
\frac{J_{2}(\uutheta_2)}{\uutheta_{2}}
\end{gather}
where ($\uutheta_1$,$\uutheta_{2}$) is the unique solution to
\begin{align}
\frac{J_{1}(\theta_1)}{\theta_1} = \frac{J_{2}(\theta_{2})}{\theta_{2}},
\end{align}
with $(\theta_1,\theta_{2})\in\Omega_{\varepsilon}$.

\textbf{\underline{Case II.c}}: If $J_{1,\tmax}\ge J_{th}(\varepsilon)$ and $J_{2,\tmax}\ge J_{th}(\varepsilon)$,
\begin{enumerate}[a)]
\item If $\theta_{1,th}= \theta_{2,th}$,
\begin{gather}
\hspace{-.5cm}R_{HARQ-IR}(\varepsilon,D_\tmax,L)=\frac{J_{th}(\varepsilon)}{\theta_{1,th}},
\end{gather}
where ($\theta_{1,th}$,$\theta_{2,th}$) is the unique solution pair to $J_1(\theta_1)=J_{th}(\varepsilon)$, and $J_{2}(\theta_{2})=J_{th}(\varepsilon)$.

\item If $\theta_{1,th}>\theta_{2,th}$, 
\begin{gather}
\hspace{-.3cm}R_{HARQ-IR}(\varepsilon,D_\tmax,L)=
\frac{J_{1}(\vvtheta_1)}{\vvtheta_{1}}
\end{gather}
where 
$\vvtheta_1$ is the smallest value of $\theta_1$ with $(\theta_1,\theta_{2})\in\Omega_{\varepsilon}$ satisfying
\begin{align}
J_1(\theta_1) = J_2(\theta_2)+J_1(\theta_1-\theta_2)
\end{align}

\item If $\theta_{1,th}<\theta_{2,th}$,
\begin{gather}
\hspace{-.5cm}R_{HARQ-IR}(\varepsilon,D_\tmax,L)=
\frac{J_{2}(\uutheta_2)}{\uutheta_{2}}
\end{gather}
where 
($\uutheta_1$,$\uutheta_{2}$) is the unique solution to
\begin{align}
\frac{J_{1}(\theta_1)}{\theta_1} = \frac{J_{2}(\theta_{2})}{\theta_{2}},
\end{align}
with $(\theta_1,\theta_{2})\in\Omega_{\varepsilon}$.
\end{enumerate}
The associated end-to-end outage probability is given by
\begin{align}
P_{\text{out}} = 1 - (1-P_{\text{out},s})(1-P_{\text{out},r}).
\end{align}
\end{Lem}
\emph{Proof}: See Appendix \ref{app:harqresult}.\hfill$\square$

\begin{Rem}
Note that due to the outage events, it is possible that certain delay constraints may not be satisfied, e.g., \textbf{Case I}.
\end{Rem}

\begin{Lem1}
The outage effective capacity of the buffer-aided diamond relay systems with HARQ-IR strategy subject to statistical delay constraints specified by $(\varepsilon,D_{\tmax})$ can be expressed as
\begin{align}
R_{HARQ-IR}(\varepsilon,D_\tmax)=\max_{L\ge0}R_{HARQ-IR}(\varepsilon,D_\tmax,L) = R_{HARQ-IR}(\varepsilon,D_\tmax,L_{\text{opt}}).
\end{align}
The associated optimal end-to-end outage probability for $L_{\text{opt}}$ is given by
\begin{align}
P_{\text{out,opt}} = 1 - (1-P_{\text{out,opt},s})(1-P_{\text{out,opt},r}).
\end{align}
\end{Lem1}

\begin{Rem}
Following the similar reasoning in Appendix \ref{app:ec}, we can show that the outage effective capacity approaches to 0 when $L\to0$ or $L\to\infty$. So $L_{\text{opt}}$ is finite, and the approach in Remark \ref{rem:optL} can be used here to derive $L_{\text{opt}}$.
\end{Rem}

\section{Numerical Results}

\begin{figure}
\begin{center}
\includegraphics[width=\figsize\textwidth]{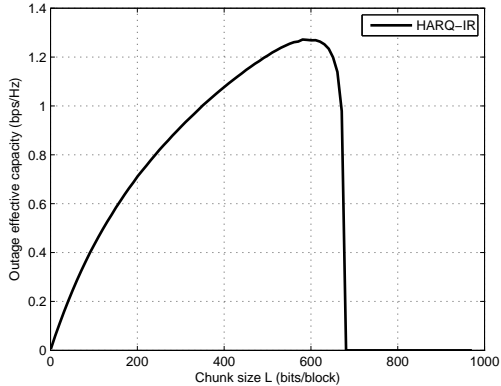}
\caption{The outage effective capacity as $L$ varies. $M=4$.}\label{fig:ecinL_dr}
\end{center}
\end{figure}

In the numerical results, we assume the fading distributions of all links follow independent Rayleigh fading with means $\E\{z_{sr_j}\} = \E\{z_{r_jd}\}=16,j=1,2$, $\tsnr=0$ dB, $T=1$ ms, $B=180$ kHz, and $D_\tmax = 1$ s. Now, $J_{1,\tmax}$ and $J_{2,\tmax}$ are finite from Remark \ref{rem:jmax}.

In Fig. \ref{fig:ecinL_dr}, we plot the outage effective capacity as a function of $L$. In this figure, we assume $\tsnr_r=5$ dB, $M=4$, and $\varepsilon=0.05$. We can see that the outage effective capacity is maximized at a finite value $L_{\text{opt}}$. Also, we can find that when $L$ is larger than certain value, the outage effective capacity vanishes immediately. This is due to the fact that when $L$ is large enough, the outage probability of each hop can be so large that the end-to-end delay constraints cannot be satisfied, i.e., \textbf{Case I} of Theorem \ref{theo:harqresult}. 

\begin{figure}
\begin{center}
\includegraphics[width=\figsize\textwidth]{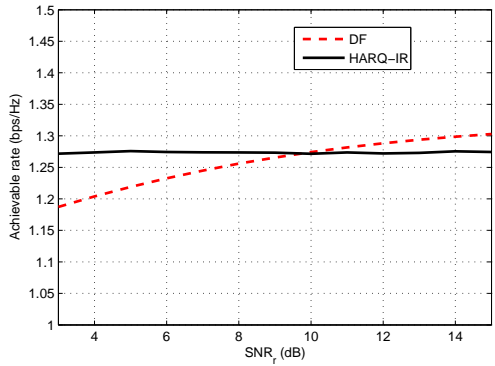}
\caption{Effective capacity as a function of $\tsnr_r$. $\varepsilon=0.05$.
}\label{fig:ecinsnr_dr}
\end{center}
\end{figure}

In Fig. \ref{fig:ecinsnr_dr}, we plot the outage effective capacity as a function of $\tsnr_r$. From the figure, it is interesting that HARQ-IR based transmission scheme can achieve larger effective capacity compared with DF protocol at relatively small SNR levels at the relays, albeit at the expense of outage. This is generally due to the fact that at smaller $\tsnr_r$ values, the effective capacity is maximized at larger $J_1(\theta_1)$, or larger $\theta_1$ equivalently. In this case, the system enjoys the benefit of average over different channel realizations provided by HARQ-IR, which can lead to larger effective capacity. On the other hand, when $\tsnr_r$ becomes large, we can see from (\ref{eq:raterdf}) that the service rate of the second hop of DF protocol increases significantly compared with the one achieved with the HARQ-IR protocol in (\ref{eq:rdrate}), which will result in much looser delay constraints $J_1(\theta_1)$ at the source, i.e., smaller $\theta_1$, and hence the effective capacity of DF protocol is larger.

\begin{figure}
\begin{center}
\includegraphics[width=\figsize\textwidth]{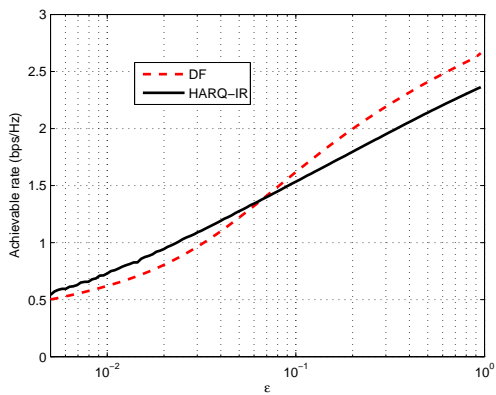}
\caption{Effective capacity as a function of $\varepsilon$. $\tsnr_r=5$ dB.}\label{fig:ecine_dr}
\end{center}
\end{figure}

\begin{figure}
\begin{center}
\includegraphics[width=\figsize\textwidth]{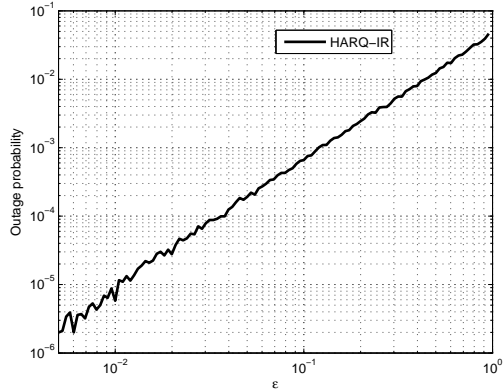}
\caption{Outage probability as a function of $\varepsilon$. $\tsnr_r=5$ dB.}\label{fig:poutine_dr}
\end{center}
\end{figure}

In Fig. \ref{fig:ecine_dr}, we plot the outage effective capacity as $\varepsilon$ varies. We assume $\tsnr_r = 5$ dB. We can find that the HARQ-IR based scheme achieves superior performance than DF protocol when $\varepsilon$ is relative small, i.e., stringent \emph{end-to-end} delay constraints. The reasoning behind is similar to previous finding. That is, at relative large $\theta_1$, the benefit provided by averaging over different channel realizations with HARQ-IR is more prominent. In Fig. \ref{fig:poutine_dr}, we plot the associated outage probability as $\varepsilon$ varies. We can find that as the delay constraints become more stringent, i.e., $\varepsilon$ decreases, the outage probability decreases. This is obvious since smaller outage probability implies less retransmissions to avoid build-up in the buffers. It is interesting that the optimal outage probability appears to be linear in the delay violation probability.

\begin{figure}
\begin{center}
\includegraphics[width=\figsize\textwidth]{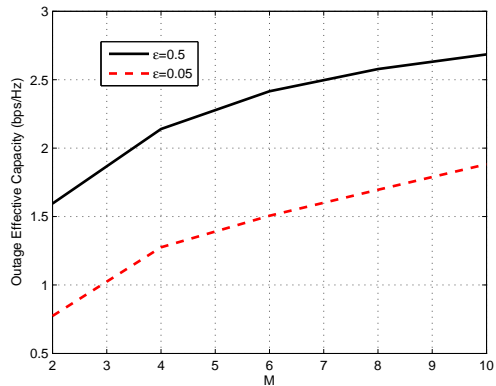}
\caption{The outage effective capacity vs. $M$. }\label{fig:ecinM_dr}
\end{center}
\end{figure}

In Fig. \ref{fig:ecinM_dr}, we plot the outage effective capacity as a function of $M$. We assume $\tsnr_r = 5$ dB and $\varepsilon=\{0.5,0.05\}$. From the figure, we can see that the outage effective capacity of the buffer-aided diamond relay systems using HARQ-IR is increasing in $M$, similar to the findings in \cite{gursoy-harq} for one-hop links.


\section{Conclusion}

In this paper, we have investigated the buffer-aided diamond relay systems with truncated HARQ-IR protocol under delay constraints. We have assumed that there is only perfect CSI at the receiver side for each link, and the transmitters send the information at a fixed rate. We have introduced the notion of \emph{outage effective capacity}, which identifies the maximum constant rate to the transmitter that can be supported by the goodput processes correctly received at the destination. We have characterized the outage effective capacity and the associated outage probability in buffer-aided diamond relay systems with HARQ-IR. Through numerical results, we have found that HARQ-IR achieves better performance than the DF protocol with perfect CSI at the transmitters as well when the $\tsnr$ at the relays are relatively small or when the delay constraints are stringent. It is interesting that the optimal end-to-end outage probability appears to be linear in the delay violation probability. 

\appendix

\subsection{Proof of Theorem \ref{theo:ec}}\label{app:ec}
Since we consider the \emph{goodput} of the departures that can be successively received at the receiver, for the state transition model in Fig. \ref{fig:state}, outage occurs when the departure in State $M-1$ cannot be correctly received at the receiver. Then, such departure processes contribute nothing to the \emph{goodput}. Note that outage occurs only at the State $M-1$, i.e., decoding failure after the $M^{\text{th}}$ transmission of the packet. We know that with fixed transmission rate $L$, the outage probability is given by
\begin{small}
\begin{align}
\hspace{-.4cm}P_{\text{out}} &= \Pr\{\text{decoding failure after the $M^{\text{th}}$ transmission of the packet}\} \nonumber\\
&=\Pr\{\text{decoding failure after the $1^{\text{st}}$ transmission of the packet}\}\nonumber\\
&\times\Pr\{\text{decoding failure after the $2^{\text{nd}}$ transmission of the packet}
\nonumber\\
&\hspace{1cm}
|\text{decoding failure after the $1^{\text{st}}$ transmission of the packet}\}\nonumber\\
&\times\cdots\times\Pr\{\text{decoding failure after the $M^{\text{th}}$ transmission of the packet}\nonumber\\
&\hspace{2cm}
|\text{decoding failure after the $(M-1)^{\text{th}}$ transmission of the packet}\}\nonumber\\
&= p_{0}\times p_{1}\times\cdots \times p_{M-1}=\prod_{m=0}^{M-1}p_m\\
& = \Pr\left\{\sum_{i=0}^{M-1}TB\log_2(1+\tsnr z_i) < L\right\}.
\end{align}
\end{small}
Obviously, $P_{\text{out}}$ varies with $L$. For the Markov model considered in (\ref{eq:stateprob}), $L$ bits of \emph{goodput} are removed from the buffer in State 0 with probability $1-P_{\text{out}}$ while 0 bit of \emph{goodput} is removed with probability $P_{\text{out}}$.

In the following, we first obtain the associated achievable outage effective capacity with given $L>0$. Regarding the Markov modulated processes, we know that \cite[Chapter 7, Example 7.2.7]{changbook}
\begin{align}
\frac{\Lambda(\theta)}{\theta} = \frac{1}{\theta}\log sp\left\{\mathbf{P}\phi(\theta)\right\}
\end{align}
where $sp\{\cdot\}$ is the spectral radius of the matrix, $\mathbf{P}$ is the state transition probability matrix (\ref{eq:stateprob}), and $\phi(\theta) =diag\{\phi_0(\theta),\ldots,\phi_{M-1}(\theta)\}$ is a diagonal matrix with each component given by the moment generating functions of the \emph{goodput} processes in $M$ states. With the above characterization of goodput processes in State 0, we have
\begin{align}
\phi_0(\theta) &= P_{\text{out}} + (1 - P_{\text{out}}) e^{\theta L}.
\end{align}
Note that 0 bit is removed in all other states. Then, we have $\phi(\theta) ={diag}\{ P_{\text{out}} + (1 - P_{\text{out}}) e^{\theta L},\underbrace{1,\ldots,1}_{M-1}\}$. We are interested in $-\frac{\Lambda(-\theta)}{\theta}$. Then, similar to \cite[Appendix A]{sami}, we can show that
\begin{align}
sp\left\{\mathbf{P}\phi(-\theta)\right\} = p_0\left(P_{\text{out}} + (1 - P_{\text{out}})e^{-\theta L}\right)y^*
\end{align}
where $y^*>0$ satisfies
\begin{align}
y^M &=  \frac{1-p_0}{p_0}y^{M-1}+\sum_{m=1}^{M-2}\frac{(1-p_m)p_{m-1}\cdots p_1}{p_0^{m} \left(P_{\text{out}} + (1 - P_{\text{out}})e^{-\theta L}\right)^m}y^{M-1-m}\nonumber\\
&+\frac{p_{M-2}\cdots p_1}{p_0^{M-1} \left(P_{\text{out}} + (1 - P_{\text{out}})e^{-\theta L}\right)^{M-1}}.
\end{align}
In addition, we can show that there is only one unique real positive root of $f(y)$ defined in (\ref{eq:fy}). In this way, we can express the achievable outage effective capacity with given $L$ as in (\ref{eq:ecfixedL}).

Then, the maximum outage effective capacity can be obtained by maximizing over the fixed transmission rate $L$. Denote the optimal solution as $L_{\text{opt}}$. We can show that there must exist a finite $L_{\text{opt}}$. Note that when $L$ is small, the outage probability approaches 0, and the outage effective capacity is approximately $\frac{L}{TB}$, which generally increases with $L$. Meanwhile as $L$ approaches infinity, we know that outage probability approaches 1, i.e., the receiver cannot correctly receive any packet, in which case the outage effective capacity becomes 0. So there must be some finite value of $L_{\text{opt}}$ that (\ref{eq:ecfixedL}) is solved, proving the results in the theorem. \hfill$\square$

\subsection{Proof of Proposition \ref{theo:eclimit}}\label{app:eclimit}

Note that with the definitions of the parameters, we can see from the discussions in Appendix \ref{app:ec} that
\begin{align}\label{eq:poutsubs}
P_{\text{out},m} =  p_{0}\times p_{1}\times\cdots \times p_{m-1}=\prod_{m=0}^{m-1}p_m.
\end{align}
By letting (\ref{eq:fy}) equal 0 and multiplying both sides of the resulting equation by $p_0^M(P_{\text{out}}+(1-P_{\text{out}})e^{-\theta L})^M$, we obtain
\begin{align}
&(p_0(P_{\text{out}}+(1-P_{\text{out}})e^{-\theta L})y)^M - (1-p_0)(P_{\text{out}}+(1-P_{\text{out}})e^{-\theta L})(p_0(P_{\text{out}}+(1-P_{\text{out}})e^{-\theta L})y)^{M-1}\nonumber\\
& -\sum_{m=1}^{M-2}(1-p_m)p_{m-1}\cdots p_{1}p_0(P_{\text{out}}+(1-P_{\text{out}})e^{-\theta L})(p_0(P_{\text{out}}+(1-P_{\text{out}})e^{-\theta L})y)^{M-m-1}\nonumber\\
&-p_{M-2}\cdots p_1p_0(P_{\text{out}}+(1-P_{\text{out}})e^{-\theta L}) = 0.
\end{align}
Combining (\ref{eq:poutsubs}) with the above equation and letting $u = p_0(P_{\text{out}}+(1-P_{\text{out}})e^{-\theta L})y$, we have
\begin{align}\label{eq:fyequiv}
&u^M- (P_{\text{out},0}-P_{\text{out},1})(P_{\text{out}}+(1-P_{\text{out}})e^{-\theta L}) u^{M-1}-\sum_{m=1}^{M-2}(P_{\text{out},m}-P_{\text{out},m+1})(P_{\text{out}}+(1-P_{\text{out}})e^{-\theta L})\nonumber\\
&\hspace{2cm}\times u^{M-1-m} - P_{\text{out},M-1}(P_{\text{out}}+(1-P_{\text{out}})e^{-\theta L}) = 0.
\end{align}
First, we can show that as $\theta\to0$, $u\to1$. We know that $(P_{\text{out}}+(1-P_{\text{out}})e^{-\theta L})\to1$ as $\theta\to0$. Then, (\ref{eq:fyequiv}) reduces to
\begin{align}
u^M- (P_{\text{out},0}-P_{\text{out},1}) u^{M-1}-\sum_{m=1}^{M-2}(P_{\text{out},m}-P_{\text{out},m+1})u^{M-1-m} - P_{\text{out},M-1} = 0.
\end{align}
It can be easily verified that $u=1$ is the unique positive solution to the above equation.

Suppose the Taylor series expansion of $u$ with respect to small $\theta$ is given by
\begin{align}\label{eq:ztaylor}
u = 1- \zeta \theta+o(\theta),
\end{align}
where $\zeta>0$ is some constant. According to (\ref{eq:ecoptL}), we can see that
\begin{align}\label{eq:ecval}
\lim_{\theta\to0}R_{\text{out}}(\tsnr,\theta) = \lim_{\theta\to0}-\frac{1}{\theta TB}\log(1-\zeta\theta+o(\theta))=\frac{\zeta}{TB}.
\end{align}
Therefore, we only need to determine the value of $\zeta$ to obtain the limit of outage effective capacity as $\theta\to0$. The Taylor series expansion of $P_{\text{out}}+(1-P_{\text{out}})e^{-\theta L}$ with respect to small $\theta$ is given by
\begin{align}\label{eq:outtaylor}
P_{\text{out}}+(1-P_{\text{out}})e^{-\theta L} = 1 - (1-P_{\text{out}})L\theta+o(\theta).
\end{align}
Substituting (\ref{eq:ztaylor}) and (\ref{eq:outtaylor}) into (\ref{eq:fyequiv}), 
and rearranging and combining the coefficients of $\theta$ gives us
\begin{align}
\left(M-(M-1)(P_{\text{out},0}-P_{\text{out},1} )-\sum_{m=1}^{M-2}(P_{\text{out},m}-P_{\text{out},m+1})(M-1-m)\right)\zeta = (1-P_{\text{out}})L,
\end{align}
which yields
\begin{align}\label{eq:zetaval}
\zeta = \frac{(1-P_{\text{out}})L}{\sum_{m=0}^{M-1}P_{\text{out},m}}.
\end{align}
Combining (\ref{eq:ecval}) and (\ref{eq:zetaval}) and replacing the parameters with the optimal values when $L=L_{\text{opt}}$ proves the result in the proposition.\hfill$\square$

\subsection{Proof of Proposition \ref{theo:t1minf}}\label{app:t1minf}

As $M\to\infty$, we know that the outage probability $P_{\text{out}}\to0$. This is obvious since as $M\to\infty$, retransmissions are triggered as long as the receiver does not receive the packet. We can rewrite (\ref{eq:fy}) as
\begin{align}\label{eq:fyrev}
f(y) &= y^M - \frac{1-p_0}{p_0}y^{M-1}
-\sum_{i=1}^{M-1}\frac{(1-p_i)p_{i-1}\cdots p_1}{p_0^{i} \left(P_{\text{out}} + (1-P_{\text{out}}) e^{-\theta L}\right)^i}y^{M-1-i} 
-\frac{\prod_{m=1}^{M-1}p_m}{p_0^{M-1} \left(P_{\text{out}} + (1-P_{\text{out}}) e^{-\theta L}\right)^{M-1}}.
\end{align}
Since the channel is modeled as independently identically distributed (IID) between frames, we have the following characterization for HARQ-T1 protocol
\begin{align}\label{eq:tp1}
p_0&=p_1=\cdots=p_{M-1}=\Pr\left\{L > TB\log_2(1+\tsnr z)\right\}
=\Pr\left\{z<\frac{2^{L/TB}-1}{\tsnr}\right\}.
\end{align}
Substituting (\ref{eq:tp1}) into (\ref{eq:fyrev}) , we have
\begin{small}
\begin{align}
f(y)&=y^M - \frac{1-p_0}{p_0}\sum_{i=0}^{M-1}\left(P_{\text{out}} + (1 - P_{\text{out}})e^{-\theta L}\right)^{-i}y^{M-1-i} 
- \frac{1}{\left(P_{\text{out}} + (1-P_{\text{out}}) e^{-\theta L}\right)^{M-1}}\nonumber\\
& = y^{M-1} \left(y - \frac{1-p_0}{p_0}\sum_{i=0}^{M-1}\left(\left(P_{\text{out}} + (1 - P_{\text{out}})e^{-\theta L}\right)y\right)^{-i} \right)
- \frac{1}{\left(P_{\text{out}} + (1-P_{\text{out}}) e^{-\theta L}\right)^{M-1}} \label{eq:harqt1-sim0}\\
& = y^{M-1} \left(y - \frac{1-p_0}{p_0}\frac{1-\left(\left(P_{\text{out}} + (1 - P_{\text{out}})e^{-\theta L}\right)y\right)^{-M}}{1- \left(\left(P_{\text{out}} + (1 - P_{\text{out}})e^{-\theta L}\right)y\right)^{-1}}\right) 
- \frac{1}{\left(P_{\text{out}} + (1-P_{\text{out}}) e^{-\theta L}\right)^{M-1}}.\label{eq:harqt1-sim1}
\end{align}
\end{small}

Letting $f(y)=0$, we have
\begin{align}\label{eq:fyder}
&y - \frac{1-p_0}{p_0}\frac{1-\left(\left(P_{\text{out}} + (1 - P_{\text{out}})e^{-\theta L}\right)y\right)^{-M}}{1- \left(\left(P_{\text{out}} + (1 - P_{\text{out}})e^{-\theta L}\right)y\right)^{-1}}
= \frac{1}{\left(\left(P_{\text{out}} + (1-P_{\text{out}}) e^{-\theta L}\right)y\right)^{M-1}}
\end{align}
We can show that $\left(P_{\text{out}} + (1 - P_{\text{out}})e^{-\theta L}\right)y>1$. Suppose that $\left(P_{\text{out}} + (1 - P_{\text{out}})e^{-\theta L}\right)y\le1$. We will have infinite value for the summation in (\ref{eq:harqt1-sim0}) as $M\to\infty$, which in turn returns $y\to\infty$ as $M\to\infty$ from (\ref{eq:fyder}), and as a result $\left(P_{\text{out}} + (1 - P_{\text{out}})e^{-\theta L}\right)y\to\infty$ as $M\to\infty$, violating the assumption.

Taking the limit of both sides of (\ref{eq:fyder}) as $M\to\infty$ and noting that $P_{\text{out}} \to0$ as $M\to\infty$, we have
\begin{align}
y - \frac{1-p_0}{p_0}\frac{1}{1- e^{\theta L}/y} = 0
\end{align}
which after rearrangement yields
\begin{align}\label{eq:yst-harqt1}
y^*= \frac{1}{p_0}\left(1-p_0+p_0e^{\theta L}\right).
\end{align}
Substituting (\ref{eq:yst-harqt1}) and $P_{\text{out}}\to0$ into (\ref{eq:ecfixedL}), we have
\begin{align}
\lim_{M\to\infty}R_{\text{out}}(\tsnr,\theta)
& = \max_{L\ge0}\left\{-\frac{1}{\theta TB}\log\left(p_0+(1-p_0)e^{-\theta L}\right)\right\}\\
& = \max_{L\ge0}\Bigg\{-\frac{1}{\theta TB}\log\bigg(1 - \Pr\left\{z>\frac{2^{L/TB}-1}{\tsnr}\right\}
\left(1-e^{-\theta L}\right)\bigg)\Bigg\},
\end{align}
proving the results in the proposition.
\hfill$\square$

\subsection{Proof of Theorem \ref{theo:harqresult}}\label{app:harqresult}

The idea of this proof follows that in \cite[Appendix D]{deli-twohopend}, except that the effective capacity limits as $\theta\to\infty$ and $\theta\to0$ are different, and the potential values of $J_1(\theta_1)$ and $J_2(\theta_2)$ lie in the range of $[0,J_{1,\tmax}]$ and $[0,J_{2,\tmax}]$, respectively. Therefore, when we iterate over all possible $(J_1,J_2)$ pairs to find the maximum constant arrival rate, we have sliced part of the $J_1-J_2$ curve characterized in Lemma \ref{lemm:J1J2relation}.

\begin{figure}
\begin{center}
\includegraphics[width=\figsize\textwidth]{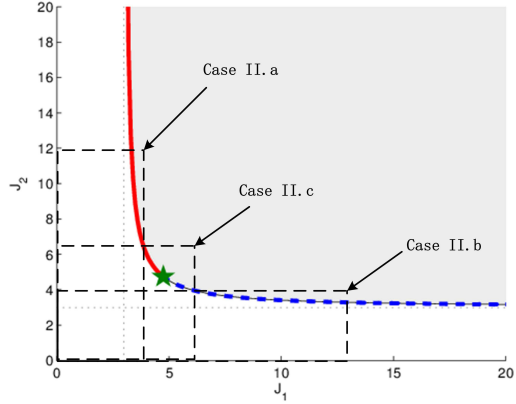}
\caption{Illustration of three cases depending on $(J_{1,\tmax},J_{2,\tmax})$. The star point denotes $(J_{th}(\varepsilon),J_{th}(\varepsilon))$.}\label{fig:harqJ1J2}
\end{center}
\end{figure}

First, we need to check if the statistical delay constraints $(D_\tmax,\varepsilon)$ can be satisfied. Substituting $J_1(\theta_1)=J_{1,\tmax}$ and $J_2(\theta_2)=J_{2,\tmax}$ into (\ref{eq:J1J2relation}), we can compare the obtained value with $\varepsilon$. If the resulting value is larger than $\varepsilon$, i.e., \textbf{Case I}, the statistical delay constraints cannot be satisfied, and hence the effective capacity is zero. If the resulting value is smaller than $\varepsilon$, we can have three different cases depending on the relationship between $(J_{1,\tmax},J_{2,\tmax})$ and $(J_{th}(\varepsilon),J_{th}(\varepsilon))$ as shown in Fig. \ref{fig:harqJ1J2}. Specifically, we have:
\begin{enumerate}[a)]
\item \textbf{Case II.a}: If $J_{1,\tmax}<J_{th}(\varepsilon)$, the intersection points of the region $[0,J_{1,\tmax}]\times [0,J_{2,\tmax}]$ with the upper boundary curve $J_2(\theta_2)=\Phi(J_1(\theta_1))$ lie in the branch with $J_2>J_1$. So we only need to iterate over this branch for potential point $(J_1,J_2)$ achieving the maximum effective capacity.
\item \textbf{Case II.b}: If $J_{2,\tmax}<J_{th}(\varepsilon)$, we only need to iterate over the branch with $J_2<J_1$ for potential point $(J_1,J_2)$ achieving the maximum effective capacity.
\item \textbf{Case II.c}: Otherwise, we need to consider the two branches jointly to identify the effective capacity.
\end{enumerate}

\underline{\textbf{Case II.a}:} Assume $J_{1,\tmax}<J_{th}(\varepsilon)$. In this case, we can relieve the statistical delay constraints at the source, i.e., decrease $J_1(\theta_1)$, or $\theta_1$ equivalently. Correspondingly, according to Lemma \ref{lemm:J1J2relation}, $J_2(\theta_2)$, and hence $\theta_2$, should increase. We can show that the queue at the relay will not affect the performance as long as $\theta_1$ and $\theta_2$ satisfies the following inequality given by
\begin{align}\label{eq:ineqcond}
J_1(\theta_1)\le J_2(\theta_2) + J_1(\theta_1-\theta_2),
\end{align}
and the effective capacity is given by
\begin{align}
R_E(\theta_1,\theta_2) = \frac{J_1(\theta_1)}{\theta_1}.
\end{align}
Note that as $J_1(\theta_1)$ increases to $J_{1,\tmax}$, $\theta_1\to\infty$, and $J_1(\theta_1-\theta_2)>0$. At the same time, $J_1(\theta_1)<J_2(\theta_2)$ for this case. The inequality (\ref{eq:ineqcond}) can be satisfied when $J_1(\theta_1)$ approaches to $J_{1,\tmax}$. On the other hand, as $J_2(\theta_2)$ increases to $J_{2,\tmax}$, $\theta_2\to\infty$, in which case $\frac{1}{\theta_1-\theta_2}J_1(\theta_1-\theta_2)$ approaches to largest possible rate of the first hop \cite{changbook}, i.e., $L$ bits/block. Then, $J_1(\theta_1-\theta_2)$ approaches minus infinity, and hence the right-hand-side of (\ref{eq:ineqcond}) is less than 0. That is, the inequality (\ref{eq:ineqcond}) cannot be satisfied when $J_2(\theta_2)$ approaches to $J_{2,\tmax}$.  Therefore, there must be a point $(\vvtheta_1,\vvtheta_2)\in\Omega_\varepsilon$ such that $\vvtheta_1$ is the smallest value of $\theta_1$ while (\ref{eq:ineqcond}) can be satisfied with equality at $(\theta_1,\theta_2)$. We can show that the effective capacity in this case is given by
\begin{align}\label{eq:caseiia}
R_{HARQ-IR}(\varepsilon,D_\tmax,L) = \sup_{(\theta_1,\theta_2)\in\Omega}R_E(\theta_1,\theta_2) = R_E(\vvtheta_1,\vvtheta_2)=\frac{J_1(\vvtheta_1)}{\vvtheta_1}.
\end{align}
Further relieving the statistical delay constraints at the source beyond $J_1(\vvtheta_1)$ will result in rate loss since the inequality (\ref{eq:ineqcond}) can not be satisfied, and the queues of the second hop will become the bottle-neck of the system.

\underline{\textbf{Case II.b}:} Assume $J_{2,\tmax}<J_{th}(\varepsilon)$. In this case, we can relieve the statistical delay constraints at the relays, i.e., decrease $J_2(\theta_2)$, or $\theta_2$ equivalently. Correspondingly, according to Lemma \ref{lemm:J1J2relation}, $J_1(\theta_1)$, and hence $\theta_1$, should increase. In this case, we know that the effective capacity is given by
\begin{align}\label{eq:rate2b}
\min\left\{\frac{J_1(\theta_1)}{\theta_1}, \frac{J_2(\theta_2)}{\theta_2}\right\}.
\end{align}
Note that as $J_1(\theta_1)$ increases to $J_{1,\tmax}$, $\theta_1\to\infty$ and hence $\frac{J_{1,\tmax}}{\theta_1}$ approaches to the minimum possible rate of the first hop, which is zero. That is, $\frac{J_2(\theta_2)}{\theta_2}>\frac{J_1(\theta_1)}{\theta_1}$ as $J_1(\theta_1)\to J_{1,\tmax}$. Similarly, $\frac{J_2(\theta_2)}{\theta_2}<\frac{J_1(\theta_1)}{\theta_1}$ as $J_2(\theta_2)\to J_{2,\tmax}$. Therefore, we can find a unique pair of $(\uutheta_1,\uutheta_2)\in\Omega_\varepsilon$ such that $\frac{J_1(\uutheta_1)}{\uutheta_1}=\frac{J_2(\uutheta_2)}{\uutheta_2}$. We can show that the effective capacity in this case is given by
\begin{align}\label{eq:caseiib}
R_{HARQ-IR}(\varepsilon,D_\tmax,L)=\sup_{(\theta_1,\theta_2)\in\Omega}R_E(\theta_1,\theta_2)=R_E(\uutheta_1,\uutheta_2)=\frac{J_2(\uutheta_2)}{\uutheta_2}.
\end{align}
Further relieving the statistical delay constraints at the relay beyond $J_2(\uutheta_2)$ will result in rate loss since the queues of the first hop will become the bottle-neck of the system.

\underline{\textbf{Case II.c}:} Assume $J_{1,\tmax}\ge J_{th}(\varepsilon)$ and $J_{2,\tmax}\ge J_{th}(\varepsilon)$. Now, we need to iterate over two branches with $J_1<J_2$ and $J_1>J_2$ to find the optimal point $(J_1,J_2)$ such that the effective capacity can be maximized. Note that this case cover the possibilities in \textbf{Case II.a} and \textbf{Case II.b}, and also the case in which symmetric delay constraints at the source and relays can achieve the maximum effective capacity, i.e., $J_1(\theta_1)=J_2(\theta_2)=J_{th}(\varepsilon)$.

The details of the derivation for the above claims in (\ref{eq:caseiia}) and (\ref{eq:caseiib}) are similar to the proof in \cite[Appendix D]{deli-twohopend}, and are omitted here. Interested readers are encouraged to find more details in \cite[Appendix D]{deli-twohopend}.

Meanwhile, we can see that the data correctly received by the destination must also be correctly received at the relays. Therefore, the outage probability of the diamond-relay channels is given by
\begin{align}
P_{\text{out}} = 1 - (1-P_{\text{out},s})(1-P_{\text{out},r}).
\end{align}



\end{document}